\documentclass[12pt,
showpacs,preprintnumbers,
amsmath,amssymb,aps,prd]{revtex4}
\usepackage{dcolumn}
\usepackage{amsthm,amscd,mathrsfs}
\usepackage{amsfonts,bm,latexsym}
\begin{document}

\title{Dirac particles' tunnelling from black rings}

\author{Qing-Quan Jiang \footnote{E-mail address: jiangqq@iopp.ccnu.edu.cn}}
\affiliation{Institute of Particle Physics, Central China Normal University,
Wuhan, Hubei 430079, People's Republic of China}

\begin{abstract}
Recent research shows that Hawking radiation can be treated as a
quantum tunnelling process, and Hawking temperature of Dirac
particles across the horizon of a black hole can be correctly
recovered via fermions tunnelling method. In this paper, motivated
by fermions tunnelling method, we attempt to apply the analysis to
derive Hawking radiation of Dirac particles via tunnelling from
black ring solutions of $5$-dimensional Einstein-Maxwell-dilaton
gravity theory. Finally, it is interesting to find as in black
hole cases, fermions tunnelling can also result in correct Hawking
temperatures for the rotating neutral, dipole and charged black
rings.
\end{abstract}


\maketitle

\section{Introduction}

Since Hawking had proved that a black hole can radiate
particles characterized by the thermal spectrum with the
temperature $T=(1/2\pi)\kappa$, where $\kappa$ is the surface
gravity of the black hole, many papers appear to correctly derive
Hawking temperature via different methods, such as gravity
collapsing method\cite{Hawking}, temperature Green
function\cite{Gibbons}, path integral\cite{HH}, Euclidean action
integral\cite{GH}, second quantum method\cite{DB}, renormalization
energy-momentum tensor\cite{BD} and more recently developed
technique called generalized tortoise coordinate
transformation(GTCT) to deal with Hawking radiation of an
evaporating black holes\cite{LZ,WC}, etc. The study of Hawking
radiation has long been attracted a lot of attentions of
theoretical physicists. The reason is partly due to the fact that
a deeper understanding of Hawking radiation may shed some lights
on seeking the underlying quantum gravity, and on the other hand,
it is the key to make the second law of thermodynamics in
spacetimes involving black holes consistent.

In recent years, a semi-classical quantum tunnelling method, first
put forward by Kraus and Wilczek\cite{KW} and then elaborated by
Parikh and Wilczek\cite{PW}, has already attracted many people's
attentions \cite{r6,r7}. Here derivation of Hawking temperature
mainly depends on the computation of the imaginary part of the
action for the classically forbidden process of s-wave emission
across the horizon. Normally, there are two approaches to obtain
the imaginary part of the action. One, first used by Parikh and
Wilczek\cite{PW} and later broadly discussed by many
papers\cite{r6,r7}, is called as the Null Geodesic method, where
the contribution to the imaginary part of the action only comes
from the integration of the radial momentum $p_r$ for the emitted
particles. The other method regards the action of the emitted
particles satisfies the relativistic Hamilton-Jacobi equation, and
solving it yields the imaginary part of the action\cite{r4}, which
is an extension of the complex path analysis proposed by
Padmanabhan et al\cite{r5}. In the two tunnelling modes, they use
the fact that the tunnelling rate for the classically forbidden
trajectory from inside to outside the horizon is given by
$\Gamma=\exp{(-\frac{2}{\hbar}\textrm{Im} I)}$, where $I$ is the
classical action of the trajectory to leading order in $\hbar$.
Where these two methods differ is in how the action is calculated.
Ref.\cite{r8} has given a detailed comparison between the
Hamilton-Jacobi ansatz and the Null Geodesic method.

Although the tunnelling method is shown very robust to
successfully derive Hawking radiation of black holes and even
black rings, most papers have only considered scalar particle's
tunnelling radiation. In fact, a black hole can radiate all types
of particles at the Hawking temperature, and the true emission
spectrum should contain contributions of both scalar particles and
fermions with all spins. Recently, applications of quantum
tunnelling methods to fermions case has first been presented in
Ref.\cite{r9} to correctly describe Hawking radiation of fermions
with spin $1/2$ via tunnelling from Rindler space-time and that
from the uncharged spherically symmetric black holes. Later, to
further prove the robustness of fermions tunnelling method, some
papers appear to discuss Hawking radiation of fermions via
tunnelling from BTZ black hole\cite{r10}, dynamical black
hole\cite{r11}, Kerr black hole\cite{r12}, Kerr-Newman black
hole\cite{r13} and more general and complicated black
holes\cite{r14}. These involved black holes share in taking $3-$
or $4-$dimensional spacetimes. For spacetimes with different
horizon topology and different dimensions, choosing a set of
appropriate $\gamma^\mu$ matrices for general covariant Dirac
equation is critical for fermions tunnelling method. In
$3$-dimensional cases, as the Pauli matrices $\sigma^i(i=1,2,3)$
behave independent each other, we can only introduce the matrices
$\sigma^i$ to act as $\gamma^\mu$ functions for the covariant
Dirac equation \cite{r10}. However for 4-dimensional spacetimes,
we need four independent matrices to well describe the matrices
$\gamma^\mu$ for the Dirac equation, and a detailed choice for the
four matrices $\gamma^\mu$ see Refs.\cite{r9,r11,r12,r13,r14}.
Then how to choose the $\gamma^\mu$ matrices for $5$-dimensional
cases?  To the best of our knowledge, five independent matrices
should be involved in our discussion. On the other hand, the
horizon topology also has an important impact on the choice for
the matrices $\gamma^\mu$\cite{r13}. In Sec.\ref{2} and \ref{3},
we will successfully introduce a set of appropriate matrices
$\gamma^\mu$ for the 5-dimensional neutral, dipole and charged
black rings with the horizon topology $S^1\times S^2$ to well
describe Dirac particles' tunnelling radiation.

Black rings in five dimensions have many unusual properties not shared by
Myers-Perry black holes with spherical topology, for instance, their event
horizon topology is $S^1\times S^2$, not spherical for the neutral, dipole
and charged black rings. (Actually, some topological black holes also have
nontrivial topology, see for example, \cite{nbmv}). Therefore, it is very
interesting to study Hawking radiation from these black ring solutions.
In Ref.\cite{r15}, scalar particles
via tunnelling from black rings has already been discussed by
using the so-called Hamilton-Jacobi method. And in \cite{r16},
following recently hot discussion on anomalous derivation of
Hawking radiation, the authors attempt to recover Hawking
temperature of black rings via gauge and gravitational anomalies
at the horizon. However, when reducing the higher dimensional
theory to the effective two dimensional theory, they also only
consider scalar field near the horizon. As far as I know, till
now, there is no references to report Hawking radiation of Dirac
particles across black rings. So it is interesting to see if
fermions tunnelling method is still applicable in such exotic
spacetime, and how to choose the matrices $\gamma^\mu$ for the
covariant Dirac equation of 5-dimensional black rings. In this
paper, we shall concentrate ourselves on Dirac particles'
tunnelling radiation from $5$-dimensional black rings via fermions
tunnelling method. We finally find as in black hole cases,
fermions tunnelling result in correct Hawking temperatures for the
rotating neutral, dipole and charged black rings.

The remainders of this paper is organized as follows. In
Sec.\ref{2}, Hawking radiation of Dirac particles via tunnelling
from the $5$-dimensional rotating neutral black ring has been
studied by improving fermions tunnelling method. Here to make a
following analysis on the rotating dipole and charged black rings
in a more unified form in Sec.\ref{3}, we deduce a general
5-dimensional metric from the rotating neutral black ring, and
discuss its Hawking radiation of Dirac particles. In fact, the
involved 5-dimensional metric is not arbitrarily taken, and after
some substitutions has a unified form as the rotating neutral,
dipole and charged black rings(see Ref.\cite{r15}). Sec.\ref{3} is
devoted to once again check the validity of fermions tunnelling
method for the rotating dipole and charged black rings.
Sec.\ref{4} contains some conclusions and discussions.

\section{Dirac particles' tunnelling from neutral black ring}\label{2}

In this section, we focus on studying Hawking radiation of Dirac
particles via tunnelling from $5$-dimensional neutral black ring.
In this paper, black rings involved are only special solutions of
the Einstein-Maxwell-Dilaton gravity model (EMD) in
$5$-dimensions, and the corresponding action takes the forms as
\begin{equation}
S=\frac{1}{16\pi }\int d^{5}x\sqrt{-g}\left(
\mathcal{R}-\frac{1}{2}\left(
\partial \phi \right)  ^{2}-\frac{1}{4}e^{-\alpha
\Phi}F^{2}\right), \label{eq1}
\end{equation}
where $F$ is a three-form field strength and $\Phi$ is a dilaton.
Black ring solutions of the action (\ref{eq1}) have special
characters: 1) they all have horizon of topology $S^1\times S^2$;
2) there exist three Killing coordinates to determine their local
symmetries; 3) there exists infinitely many different black rings
solutions carrying the same mass, angular momentum and electric
charge, etc. In this paper, the rotating neutral, dipole and
charged black rings accompanied by the action (\ref{eq1}) are
involved in our discussion. First, we consider the case of the
$5$-dimensional neutral black ring. The neutral black ring in
$5$-dimensional EMD theory has been given by\cite{RE}
\begin{eqnarray}
 ds^{2}  &  =&-\frac{F(y)}{F(x)}\left(
dt-C(\nu,\lambda)R\frac{1+y}{F(y)}
d\psi \right)  ^{2} \nonumber\\
&  & +\frac{R^{2}}{(x-y)^{2}}F(x)\left[
-\frac{G(y)}{F(y)}d\psi^{2}
-\frac{dy^{2}}{G(y)}+\frac{dx^{2}}{G(x)}+\frac{G(x)}{F(x)}d\varphi^{2}\right]
,\label{eq2}
\end{eqnarray}
where
\begin{eqnarray*}
&& F(\xi) =1+\lambda \xi,\quad G(\xi)=(1-\xi^{2})(1+\nu \xi),\\
&& C(\nu,\lambda)
=\sqrt{\lambda(\lambda-\nu)\frac{1+\lambda}{1-\lambda}}.
\end{eqnarray*}
The parameters $\lambda$ and $\nu$ are dimensionless and takes
values in the range $(0<\nu\leq\lambda<1)$, and to avoid the
conical singularity also at $x = 1$, $\lambda$ and $\nu$ must be
related to each other via $\lambda=2\nu/(1+\nu^2)$. The coordinate
$\phi$ and $\psi$ are two cycles of the black ring, and $x$ and
$y$ takes the range as $-1\leq x \leq 1$ and $-\infty \leq y \leq
-1$. The constant $R$ has the dimensional of length and for thin
large rings corresponds roughly to the radius of the ring
circle\cite{EZ}. The horizon is located at $y=y_h=-1/\nu$. The
mass of the black ring is $M=3\pi R^2 \lambda/[4(1-\nu)]$, and its
angular momentum takes $J=\pi R^3
\sqrt{\lambda(\lambda-\nu)(1+\lambda)}/[2(1-\nu)^2]$. In addition,
the spacetime contains three Killing coordinates $t$, $\varphi$
and $\psi$. Next, we shall study Dirac particles' tunnelling from
the above neutral black ring. For simplicity, we take
\begin{eqnarray}
&& \mathcal{M}(x,y) =\frac{F(y)}{F(x)}\left(
1-\frac{C^2(\nu,\lambda)(1+y)^{2}(x-y)^2}
{F^2(x)G(y)+C^2(\nu,\lambda)(1+y)^{2}(x-y)^2}\right) , \nonumber\\
&& \mathcal{N}(x,y)  = -\left(
\frac{R^{2}}{(x-y)^{2}}\frac{F(x)}{G(y)}\right) ^{-1}, \nonumber\\
&& N^\psi(x,y)=-\frac{C(\nu,\lambda)R(1+y)F(y)(x-y)^2}{C^2(\nu,\lambda)%
(x-y)^2R^2(1+y)^2+R^2F^2(x)G(y)}, \nonumber\\
&& g_{\psi\psi}(x,y)=-\frac{C^2(\nu,\lambda)(x-y)^2
R^2(1+y)^2+R^2F^2(x)G(y)}{F(x)F(y)(x-y)^2}, \nonumber\\
&& g_{xx}(x,y)=\frac{R^2F(x)}{(x-y)^2G(x)}, \quad g_{\varphi
\varphi}(x,y)=\frac{R^2G(x)}{(x-y)^2}. \label{eq3}
\end{eqnarray}
Now the new form of neutral black ring (\ref{eq2}) changes as
\begin{eqnarray}
ds^{2}&=&-\mathcal{M}(x,y) dt^{2}+\frac{1}{\mathcal{N}(x,y)}dy^{2} \nonumber\\
&+& g_{\psi \psi}(x,y)\Big(d\psi + N^{\psi}(x,y)
dt\Big)^{2}+g_{xx}(x,y)dx^{2}+g_{\varphi
\varphi}(x,y)d\varphi^{2}. \label{ds}
\end{eqnarray}
At the event horizon of the neutral black ring, the coefficients
in Eq.(\ref{eq3}) obviously obey
\begin{equation}
\mathcal{M}(x,y_h)=\mathcal N (x,y_h)=0, \quad
N^\psi(x,y_h)=-\Omega_h, \label{eq5}
\end{equation}
where $y=y_h$ is the event horizon of the neutral black ring and
$\Omega_h$ is the angular velocity of the black ring at the event
horizon. Throughout this paper, the $5$-dimensional spacetime
coordinates are always chosen as $x^{\mu}=(t, y, \varphi, x,
\psi)$.

Now we focus on studying Dirac particles' tunnelling from the
rotating neutral black ring. In curved spacetime, Dirac particles'
motion equation satisfies the following covariant Dirac equation
\begin{equation}
i\gamma ^a e_a^\mu D_\mu \Psi -\frac{m}{\hbar }\Psi = 0,
\label{eq6}
\end{equation}
where $D_\mu$ is the spinor covariant derivative defined by
$D_\mu=\partial_\mu+\frac{1}{4}\omega_\mu^{ab}\gamma_{[a}\gamma_{b]}$,
and $\omega_\mu^{ab}$ is the spin connection corresponding to the
tetrad $e_a^\mu$. In this paper, we choose the matrices
$\gamma^a=(\gamma^0, \gamma^3, \gamma^4, \gamma^1, \gamma^2)$ for
the $5$-dimensional rotating neutral black ring, where
\begin{eqnarray}
 &&\gamma ^0 = \left(
{{\begin{array}{*{20}c}
 0 \hfill & I \hfill \\
 -I \hfill & 0 \hfill \\
\end{array} }} \right),
\quad \gamma ^1 = \left(
{{\begin{array}{*{20}c}
 0 \hfill & {\sigma ^1} \hfill \\
 {\sigma ^1} \hfill & 0 \hfill \\
\end{array} }} \right),
\quad \gamma ^2 =  \left( {{\begin{array}{*{20}c}
 0 \hfill & {\sigma ^2} \hfill \\
 {\sigma ^2} \hfill & 0 \hfill \\
\end{array} }} \right), \nonumber\\
&& \gamma ^3 = \left( {{\begin{array}{*{20}c}
 0 \hfill & {\sigma ^3} \hfill \\
 {\sigma ^3} \hfill & 0 \hfill \\
\end{array} }} \right),
\quad \gamma ^4 = \left( {{\begin{array}{*{20}c}
 -I \hfill & 0 \hfill \\
 0 \hfill & I \hfill \\
\end{array} }} \right), \label{eq7}
\end{eqnarray}
and the $\sigma^i(i=1,2,3)$ are the Pauli matrices, which are
given by
\begin{equation}
\sigma^1= \left( {{\begin{array}{*{20}c}
 0 \hfill & 1 \hfill \\
 1 \hfill & 0 \hfill \\
\end{array} }} \right),
\quad \sigma^2= \left( {{\begin{array}{*{20}c}
 0 \hfill & -i \hfill \\
 i \hfill & 0 \hfill \\
\end{array} }} \right),
\quad  \sigma^3= \left( {{\begin{array}{*{20}c}
 1 \hfill & 0 \hfill \\
 0 \hfill & -1 \hfill \\
\end{array} }} \right).\label{eq8}
\end{equation}
According to the new form of the rotating neutral black ring
(\ref{ds}), the tetrad field $e_a^\mu$ can be constructed as
\begin{eqnarray}
e_0^\mu &=&\Big(\frac{1}{\sqrt{\mathcal{M}}},0,0,0,-\frac{N^\psi}{\sqrt{\mathcal{M}}}\Big),\nonumber\\
e_1^\mu &=&\Big(0,\sqrt{\mathcal{N}},0,0,0\Big), \nonumber\\
e_2^\mu
&=&\Big(0,0,\frac{1}{\sqrt{g_{\varphi\varphi}}},0,0\Big), \nonumber\\
e_3^\mu &=&\Big(0,0,0,\frac{1}{\sqrt{g_{xx}}},0\Big), \nonumber\\
 e_4^\mu
&=&\Big(0,0,0,0,\frac{1}{\sqrt{g_{\psi\psi}}}\Big). \label{eq10}
\end{eqnarray}

As Dirac particles taking spin ${1}/{2}$, when measuring spin
along $y$ direction, there would be two cases. One is spin up
case, which shares the same direction as $y$, and the other(spin
down) case takes the opposite direction. In the Pauli matrix
$\sigma^3$ representation, they can explicitly expressed by the
eigenvectors $\xi_{\uparrow/\downarrow }$, and the corresponding
eigenvalues are $1/-1$. In this paper, we only refer to spin field
for the upper case $(\xi_\uparrow)$. In fact, after a completely
same step for spin down $(\xi_\downarrow)$ case, we can also get
the same result. We employ the following ansatz for Dirac field
with spin up case as
\begin{eqnarray}
 \Psi_{\uparrow}(t, y, \varphi, x, \psi)&=&
 \left( {{\begin{array}{*{20}c}
 A(t, y, \varphi, x, \psi)\xi_\uparrow \hfill \\
 B(t, y, \varphi, x, \psi)\xi_\uparrow \hfill \\
\end{array} }} \right)\exp\Big[\frac{i}{\hbar}I_\uparrow(t, y, \varphi, x,
\psi)\Big] \nonumber\\
&=&\left( {{\begin{array}{*{20}c}
A(t, y, \varphi, x, \psi) \hfill \\
~~~~~~~~~0 \hfill \\
B(t, y, \varphi, x, \psi)\hfill \\
~~~~~~~~~0 \hfill \\
\end{array} }} \right)\exp\Big[\frac{i}{\hbar}I_\uparrow(t, y, \varphi, x,
\psi)\Big]. \label{eq9}
\end{eqnarray}

Substituting the above ansatz (\ref{eq9}) for upper-spinning state into the
covariant Dirac equation (\ref{eq6}), then applying WKB approximation and
keeping the prominent terms, we can get the following equations
\begin{eqnarray}
B\Big(\frac{1}{\sqrt{\mathcal{M}}}\partial_t
I_\uparrow+\sqrt{\mathcal{N}}\partial_yI_\uparrow-\frac{N^\psi}{\sqrt{\mathcal{M}}}\partial_\psi
I_\uparrow \Big)
+A\Big(m-\frac{1}{\sqrt{g_{\varphi\varphi}}}\partial_\varphi
I_\uparrow \Big)&=& 0,\label{eq11} \\
 B\Big(\frac{1}{\sqrt{g_{xx}}}\partial_x
I_\uparrow+\frac{i}{\sqrt{g_{\psi\psi}}}\partial_\psi I_\uparrow
\Big)&=& 0,\label{eq12} \\
 A\Big(\frac{1}{\sqrt{\mathcal{M}}}\partial_t
I_\uparrow-\sqrt{\mathcal{N}}\partial_yI_\uparrow-\frac{N^\psi}{\sqrt{\mathcal{M}}}\partial_\psi
I_\uparrow
\Big)-B\Big(m+\frac{1}{\sqrt{g_{\varphi\varphi}}}\partial_\varphi
I_\uparrow \Big)&=& 0, \label{eq13}\\
  A\Big(\frac{1}{\sqrt{g_{xx}}}\partial_x
I_\uparrow+\frac{i}{\sqrt{g_{\psi\psi}}}\partial_\psi I_\uparrow
\Big)&=& 0.\label{eq14}
\end{eqnarray}
In fact, the derivatives of $A$ and $B$, and the components
$\frac{1}{4}\omega_\mu^{ab}\gamma_{[a}\gamma_{b]}$ are all of
order $\mathcal{O}(\hbar)$, and according to WKB approximation
have already been neglected for the above equations. Considering
the symmetries of the rotating neutral black ring, we employ the
following ansatz
\begin{equation}
I_\uparrow=-\mathcal{E} t+ \mathcal{J}\psi+\mathcal{L}\varphi
+\mathcal{W}(x,y)+\mathcal{K},\label{eq15}
\end{equation}
where $\mathcal{E}$, $\mathcal{J}$ and $\mathcal{L}$ are all real
constants which respectively represent the emitted particle's
energy and angular momentum corresponding to the angles $\psi$ and
$\varphi$ , and $\mathcal{K}$ is a complex constant(where we
consider only the positive frequency contributions without loss of
generality). Inserting the ansatz (\ref{eq15}) into
Eqs.(\ref{eq11})(\ref{eq12}),(\ref{eq13}),(\ref{eq14}), and
expanding the resulting equations near the event horizon of the
black ring, we have
\begin{eqnarray}
 B\Big(\frac{-\mathcal{E}+\Omega_h
\mathcal{J}}{\sqrt{\mathcal{M}_{,y}(x,y_h)(y-y_h)}}
+\sqrt{\mathcal{N}_{,y}(x,y_h)(y-y_h)}\partial_y \mathcal{W}(x,y)
\Big)&&
\nonumber\\
+A\Big(m-\frac{\mathcal{L}}{\sqrt{g_{\varphi\varphi}(x,y_h)}}
\Big)= 0,&&
\label{eq16} \\
 B\Big(\frac{1}{\sqrt{g_{xx}(x,y_h)}}\partial_x \mathcal{W}(x,y)+\frac{i}{\sqrt{g_{\psi\psi}(x,y_h)}}\mathcal{J}
\Big)= 0, && \label{eq17} \\
A\Big(\frac{-\mathcal{E}+\Omega_h
\mathcal{J}}{\sqrt{\mathcal{M}_{,y}(x,y_h)(y-y_h)}}
-\sqrt{\mathcal{N}_{,y}(x,y_h)(y-y_h)}\partial_y
\mathcal{W}(x,y)\Big) && \nonumber\\
-B\Big(m+\frac{\mathcal{L}}{\sqrt{g_{\varphi\varphi}(x,y_h)}}
\Big)= 0,&& \label{eq18}  \\
A\Big(\frac{1}{\sqrt{g_{xx}(x,y_h)}}\partial_x
\mathcal{W}(x,y)+\frac{i}{\sqrt{g_{\psi\psi}(x,y_h)}}\mathcal{J}
\Big)= 0. && \label{eq19}
\end{eqnarray}
Here $\mathcal{M}_{,y}(x,y_h)=\partial_y
\mathcal{M}(x,y)|_{y=y_h}$  and
$\mathcal{N}_{,y}(x,y_h)=\partial_y \mathcal{N}(x,y)|_{y=y_h}$.
Now we carry on an explicit analysis on the above equations. From
Eqs.(\ref{eq17}) and (\ref{eq19}) can we obtain
\begin{equation}
\partial_x
\mathcal{W}(x,y)=-i\sqrt{\frac{g_{xx}(x,y_h)}{g_{\psi\psi}(x,y_h)}}\mathcal{J}.
\label{eq20}
\end{equation}
And from Eqs.(\ref{eq16}) and (\ref{eq18}), one can easily see the
two equations have a non-trivial solution for $A$ and $B$ if and
only if the determinant of the coefficient matrix vanishes, so we
have
\begin{equation}
\partial_y \mathcal{W}(x,y)=\pm \frac{\sqrt{\Big(\mathcal{E}-\Omega_h\mathcal{J}
\Big)^2+\mathcal{M}_{,y}(x,y_h)(y-y_h)\Big(m^2-\frac{\mathcal{L}^2}{g_{\varphi\varphi}}\Big)}}
{\sqrt{\mathcal{M}_{,y}(x,y_h)\mathcal{N}_{,y}(x,y_h)}(y-y_h)}.
\label{eq21}
\end{equation}
It should be noted that Eq.(\ref{eq20}) implies near the horizon
of the black ring $\partial_x \mathcal{W}(x,y)$ has no explicit
$y$ dependence. On the other hand, in Eq.(\ref{eq21})
$\mathcal{M}_{,y}(x,y_h)$ and $\mathcal{N}_{,y}(x,y_h)$ are both
related to the coordinate $x$, but their product
$\mathcal{M}_{,y}(x,y_h)\cdot\mathcal{N}_{,y}(x,y_h)$ is
independent of $x$. So, near the horizon ($y\simeq y_h$),
$\partial_y \mathcal{W}(x,y)$ is independent of $x$. Now the
function $\mathcal{W}(x,y)$ can be separated as
$\mathcal{W}(x,y)=\mathcal{W}(x)+\mathcal{W}(y)$, which means near
the horizon of the black ring
$\partial_x\mathcal{W}(x,y)=\partial_x\mathcal{W}(x)$ and
$\partial_y\mathcal{W}(x,y)=\partial_y\mathcal{W}(y)$.

The WKB approximation tells us the tunnelling rate for the
classically forbidden trajectory from inside to outside the
horizon is related to the imaginary part of the emitted particle's
action across the event horizon. Now our first job is to find the
imaginary part of the action. From Eq.(\ref{eq15}), we find only
$\mathcal{W}(x,y) $ and $\mathcal{K}$ yield contributions to the
imaginary part of the action. As $\mathcal{K}$ is a complex
constant, all is focus on computing $\mathcal{W}(x) $ and
$\mathcal{W}(y)$. In fact, after an integration on
Eq.(\ref{eq20}), $\mathcal{W}(x)$ must be given by a complex
constant, so will yield a contribution to the imaginary part of
the action. From Eq.(\ref{eq21}) yields
\begin{equation}
\mathcal{W}_\pm (y)=\pm i\pi
\frac{\mathcal{E}-\Omega_h\mathcal{J}}{\sqrt{\mathcal{M}_{,y}(x,y_h)\mathcal{N}_{,y}(x,y_h)}},
\end{equation}
where $+/-$ sign corresponds to outgoing/incoming solutions. As we
all know, the tunnelling probabilities is proportional to the
imaginary part of the action. So when particles tunnel across the
horizon each way, the outgoing and ingoing rates are respectively
given by
\begin{eqnarray}
&& P_{out}=\exp\Big[-\frac{2}{\hbar} \textrm{Im}
I_\uparrow\Big]=\exp\Big[-\frac{2}{\hbar}\Big(\textrm{Im}
\mathcal{W}_+(y)+ \textrm{Im} \mathcal{W}(x)+\textrm{Im}
\mathcal{K}\Big)\Big],\nonumber\\
&& P_{in}=\exp\Big[-\frac{2}{\hbar} \textrm{Im}
I_\uparrow\Big]=\exp\Big[-\frac{2}{\hbar}\Big(\textrm{Im}
\mathcal{W}_-(y)+ \textrm{Im} \mathcal{W}(x)+\textrm{Im}
\mathcal{K}\Big)\Big].
\end{eqnarray}
Noted that any particles classically enter the horizon with no
barrier, that means the tunnelling rate should be unity for
incoming particles crossing the horizon. In our case, that implies
$\textrm{Im} \mathcal{W}_-(y)=- \textrm{Im}
\mathcal{W}(x)-\textrm{Im} \mathcal{K}$. Set $\hbar$ to unity, and
the tunnelling probability of Dirac particles crossing from inside
to outside horizon is naturally written as
\begin{eqnarray}
\Gamma & =&\exp{\Big[-4\textrm{Im}\mathcal{W}_+(y)}\Big]\nonumber\\
&=&\exp{\Big[-
\frac{4\pi}{\sqrt{\mathcal{M}_{,y}(x,y_h)\mathcal{N}_{,y}(x,y_h)}}}
\Big(\mathcal{E}-\Omega_h\mathcal{J}\Big)\Big], \label{eq24}
\end{eqnarray}
which results in the expected temperature of the rotating neutral
black ring
\begin{eqnarray}
T=\frac{\sqrt{\mathcal{M}_{,y}(x,y_h)\mathcal{N}_{,y}(x,y_h)}}{4\pi}
=\frac{1}{4\pi R}\frac{1+\nu}{\sqrt{\nu}}\sqrt{\frac{1-\lambda
}{\lambda(1+\lambda)}}. \label{eq25}
\end{eqnarray}

This result is exactly consistent with that in
Refs.\cite{r15,r16,RE}, where respectively present the correct
Hawking temperature of the rotating neutral black ring by using
the so-called Hamilton-Jacobi method,  anomalous cancellation
method and the original definition of the surface gravity. Noted
that the resulting temperature (\ref{eq25}) is only for Dirac
particles with spin up. For spin down case, taking a manner fully
analogous to the spin up case will result the same result, which
means both spin up and spin down particles are emitted at the same
rate. So such treatment does not loss the generality of fermions
tunnelling method. In addition, the tunnelling rate (\ref{eq24})
is derived by neglecting the higher terms about $\mathcal{E}$ and
$\mathcal{J}$, and the resulting spectrum is purely thermal. If we
consider energy and angular momentum conservation when particles
tunnelling out from the horizon, the higher terms will be present
in the tunnelling rate, and the radiation spectrum is not thermal,
and related to the change of Bekenstein-Hawking entropy, which
discussed a lot in Refs.\cite{PW,r6,r7,r15}. In the next section,
to further verify the validity of application of fermions
tunnelling method to black rings, we additionally take dipole and
charged black ring as an example to discuss their Hawking
radiation of Dirac particles.

\section{Dirac particles' tunnelling from dipole and charged black
rings}\label{3}

In the section, we will discuss Hawking radiation of Dirac
particles via tunnelling from dipole and charged black rings, and
expect to result in correct Hawking temperatures.

\subsection{Dipole black ring}

Dipole black ring shares the same action (\ref{eq1}) as neutral
black ring, so they physically take many similar characters. The
five dimensional dipole black rings was first constructed in
\cite{RE}, its metric takes the form as
\begin{eqnarray}
ds^{2} &=&-\frac{F(y)}{F(x)}\left(  \frac{H(x)}{H(y)}\right)
^{N/3}\left( dt-C(\nu,\lambda)R\frac{1+y}{F(y)}d\psi \right)
^{2}\nonumber\\
 & & +\frac{R^{2}}{(x-y)^{2}}F(x)
\left(  H(x)H^{2}(y)\right)  ^{N/3}\nonumber\\
&  &\times \left[
-\frac{G(y)}{F(y)H^{N}(y)}d\psi^{2}-\frac{dy^{2}}{G(y)}+\frac{dx^{2%
}}{G(x)}+\frac{G(x)}{F(x)H^{N}(x)}d\varphi^{2}\right] ,
\label{eq26}
\end{eqnarray}
where $F(\xi), G(\xi)$ and $C(\nu,\lambda)$ are of the same form
as neutral black ring, and $H(\xi) =1-\mu \xi$ ($0\leq\mu<1$). The
dilaton coupling constant is related to the dimensionless constant
$N$ as $\alpha^2=(\frac{4}{N}-\frac{4}{3})(0<N\leq 3)$. The
horizon is also located at $y=y_H=-1/\nu$. Taking the limit of
$\mu=0$ in Eq.(\ref{eq26}), this solution degenerates into neutral
black ring\cite{RE}. In suitable limits, dipole black ring also
contains Myers-Perry black hole\cite{MP}. This metric (\ref{eq26})
takes the same form as (\ref{eq2}), so we can apply the same
procedure in Sec.(\ref{2}) to correctly recover Hawking
temperature of dipole black ring. Before that, we take
\begin{eqnarray}
&& \mathcal{M}(x,y)
=\frac{F(y)}{F(x)}\left(\frac{H(x)}{H(y)}\right)^{N/3}\left(
1-\frac{C^2(\nu,\lambda)(1+y)^{2}(x-y)^2}
{F^2(x)G(y)+C^2(\nu,\lambda)(1+y)^{2}(x-y)^2}\right) , \nonumber\\
&& \mathcal{N}(x,y)  = -\left(
\frac{R^{2}}{(x-y)^{2}}\frac{F(x)}{G(y)}\left(H(x)H^2(y)\right)^{N/3}\right) ^{-1}, \nonumber\\
&& N^\psi(x,y)=-\frac{C(\nu,\lambda)R(1+y)F(y)(x-y)^2}{C^2(\nu,\lambda)%
(x-y)^2R^2(1+y)^2+R^2F^2(x)G(y)}, \nonumber\\
&& g_{\psi\psi}(x,y)=-\frac{C^2(\nu,\lambda)(x-y)^2
R^2(1+y)^2+R^2F^2(x)G(y)}{F(x)F(y)(x-y)^2}\left(\frac{H(x)}{H(y)}\right)^{N/3}, \nonumber\\
&& g_{xx}(x,y)=\frac{R^2F(x)}{(x-y)^2G(x)}\left(H(x)H^2(y)\right)^{N/3}, \nonumber\\
&& g_{\varphi
\varphi}(x,y)=\frac{R^2G(x)}{(x-y)^2}\frac{\left(H(x)H^2(y)\right)^{N/3}}{H^N(x)},
\label{eq27}
\end{eqnarray}
which results in the metric (\ref{eq26}) taking the same form as
(\ref{ds}). At the horizon, the functions $\mathcal{M}(x,y),
\mathcal{N}(x,y)$ and $N^\psi(x,y)$ still satisfy Eq.(\ref{eq5}).
Now substituting the matrices $\gamma^a$ (\ref{eq7}) and the
tetrad $e_a^\mu$ (\ref{eq10}) into the covariant Dirac equation
(\ref{eq6}), and then adopting the same procedure present in
Sec.(\ref{2}), one can reads out Hawking temperature of Dirac
particles via tunnelling from dipole black ring
\begin{eqnarray}
T=\frac{\sqrt{\mathcal{M}_{,y}(x,y_h)\mathcal{N}_{,y}(x,y_h)}}{4\pi}
=\frac{1}{4\pi
R}\frac{\nu^{(N-1)/2}(1+\nu)}{(\mu+\nu)^{N/2}}\sqrt{\frac{1-\lambda
}{\lambda(1+\lambda)}}.
\end{eqnarray}
This result has been identically derived by using Hamilton-Jacobi
method \cite{r15} and anomalous cancellation method \cite{r16},
where particles across the horizon are only for scalar cases. Note
that dipole black ring actually contains a gauge field. Here we do
not consider its effect because it is magnetic, and its electric
dual are two-form fields that do not couple to point particles(see
Chen and He's paper in \cite{r16}). In the next subsection, we
will further study Hawking radiation of a rotating black ring with
a single electric charge by using fermions tunnelling method.

\subsection{Charged black ring}

In this subsection, we consider Hawking radiation from black ring
with only one single electric charge \cite{E}. For black rings
with two or three charges \cite{EE}, we can take the similar
procedure to get the correct results. The metric of black ring
with a single electric charge can be written in the form
consistent with neutral and dipole cases as
\begin{eqnarray}
ds^{2} &=&-\frac{F(y)}{F(x)K^2(x,y)}\left(
dt-C(\nu,\lambda)R\frac{1+y}{F(y)}\cosh^2\alpha d\psi \right)
^{2}\nonumber\\
 & & +\frac{R^{2}}{(x-y)^{2}}F(x)
\left[-\frac{G(y)}{F(y)}d\psi^{2}-\frac{dy^{2}}{G(y)}+\frac{dx^{2%
}}{G(x)}+\frac{G(x)}{F(x)}d\varphi^{2}\right] , \label{eq29}
\end{eqnarray}
where some tricks are needed to reduce the original metric of
black ring with a single electric charge to the form as
(\ref{eq29}) (refer to Chen and He's paper in \cite{r16}). Here
$F(\xi)$ and $G(\xi)$ are defined as before, and
$K(x,y)=1+\lambda(x-y)\sinh^2\alpha/F(x)$, where $\alpha$ is the
parameter representing the electric charge. The metric also has a
killing horizon at $y=y_h=-1/\nu$. The dilaton field is
$e^{-\Phi}=K(x,y)$, and the gauge fields accompanied by the metric
are
\begin{equation}
\mathcal{A}_t=\frac{\lambda(x-y)\sinh\alpha\cosh\alpha}{F(x)K(x,y)},
\quad
\mathcal{A}_\psi=\frac{C(\nu,\lambda)R(1+y)\sinh\alpha\cosh\alpha}{F(x)K(x,y)},\label{eq30}%
\end{equation}
with the electric charge
$Q=2M\sinh2\alpha/\left(3\left(1+\frac{4}{3}\sinh^2\alpha\right)\right)$.
To do an explicit computation on Hawking radiation of the black
ring, we first introduce the following substitution
\begin{eqnarray}
&& \mathcal{M}(x,y) =\frac{F(y)}{F(x)K^2(x,y)}\left(
1-\frac{C^2(\nu,\lambda)(1+y)^{2}(x-y)^2\cosh^4\alpha}
{F^2(x)G(y)K^2(x,y)+C^2(\nu,\lambda)(1+y)^{2}(x-y)^2\cosh^4\alpha}\right) , \nonumber\\
&& \mathcal{N}(x,y)  = -\left(
\frac{R^{2}}{(x-y)^{2}}\frac{F(x)}{G(y)}\right) ^{-1}, \nonumber\\
&& N^\psi(x,y)=-\frac{C(\nu,\lambda)R(1+y)F(y)(x-y)^2\cosh^2\alpha}{C^2(\nu,\lambda)%
(x-y)^2R^2(1+y)^2\cosh^4\alpha+R^2F^2(x)G(y)K^2(x,y)}, \nonumber\\
&& g_{\psi\psi}(x,y)=-\frac{C^2(\nu,\lambda)(x-y)^2
R^2(1+y)^2\cosh^4\alpha+R^2F^2(x)G(y)K^2(x,y)}{F(x)F(y)(x-y)^2K^2(x,y)}, \nonumber\\
&& g_{xx}(x,y)=\frac{R^2F(x)}{(x-y)^2G(x)}, \quad  g_{\varphi\varphi}(x,y)=\frac{R^2G(x)}{(x-y)^2}, \label{eq31}%
\end{eqnarray}
where at the event horizon $\mathcal{M}(x,y), \mathcal{N}(x,y)$
and $N^\psi(x,y)$ take the values in Eq.(\ref{eq5}). Now the
metric (\ref{eq29}) has the same form as (\ref{ds}). In the
spacetime, gauge fields (\ref{eq30}) couple to Dirac particles, so
we should introduce the following covariant Dirac equation
\begin{equation}
i\gamma ^a e_a^\mu \left(D_\mu+\frac{ie}{\hbar}\mathcal{A}_\mu
\right) \Psi -\frac{m}{\hbar }\Psi = 0. \label{32}
\end{equation}
Taking the same matrices $\gamma^a$ and tetrad fields $e_a^\mu$ as
those in Eqs.(\ref{eq7}) and (\ref{eq10}) for the black ring, and
employing the ansatz (\ref{eq9}) for the spin-up Dirac particles
and then expanding the resulting equation near the horizon yields
\begin{eqnarray}
 B\Big(\frac{-\mathcal{E}+\Omega_h
\mathcal{J}+e\Phi_h}{\sqrt{\mathcal{M}_{,y}(x,y_h)(y-y_h)}}
+\sqrt{\mathcal{N}_{,y}(x,y_h)(y-y_h)}\partial_y \mathcal{W}(x,y)
\Big)&&
\nonumber\\
+A\Big(m-\frac{\mathcal{L}}{\sqrt{g_{\varphi\varphi}(x,y_h)}}
\Big)= 0,&&
\label{eq33} \\
 B\Big(\frac{1}{\sqrt{g_{xx}(x,y_h)}}\partial_x
 \mathcal{W}(x,y)+\frac{i}{\sqrt{g_{\psi\psi}(x,y_h)}}(\mathcal{J}+\mathcal{A}_\psi(x,y_h))
\Big)= 0, && \label{eq34} \\
A\Big(\frac{-\mathcal{E}+\Omega_h
\mathcal{J}+e\Phi_h}{\sqrt{\mathcal{M}_{,y}(x,y_h)(y-y_h)}}
-\sqrt{\mathcal{N}_{,y}(x,y_h)(y-y_h)}\partial_y \mathcal{W}(x,y)
\Big) && \nonumber\\
-B\Big(m+\frac{\mathcal{L}}{\sqrt{g_{\varphi\varphi}(x,y_h)}}
\Big)= 0,&& \label{eq35}  \\
A\Big(\frac{1}{\sqrt{g_{xx}(x,y_h)}}\partial_x
 \mathcal{W}(x,y)+\frac{i}{\sqrt{g_{\psi\psi}(x,y_h)}}(\mathcal{J}+\mathcal{A}_\psi(x,y_h))
\Big)= 0, && \label{eq36}
\end{eqnarray}
where
$\Phi_h=\mathcal{A}_t(x,y_h)+\Omega_h\mathcal{A}_\psi(x,y_h)$ is
the electric chemical potential at the horizon and $\Omega_h$ is
the angular velocity at the horizon. Carrying on a similar
analysis of the neutral black ring, we easily find the tunnelling
rate of charged Dirac particles across the horizon of the charged
black ring taking the form as
\begin{equation}
\Gamma  =\exp{\Big[-
\frac{4\pi}{\sqrt{\mathcal{M}_{,y}(x,y_h)\mathcal{N}_{,y}(x,y_h)}}}
\Big(\mathcal{E}-\Omega_h\mathcal{J}-e\Phi_h\Big)\Big].
\label{eq37}
\end{equation}
The Hawking temperature of the charged black ring is then given by
\begin{eqnarray}
T=\frac{\sqrt{\mathcal{M}_{,y}(x,y_h)\mathcal{N}_{,y}(x,y_h)}}{4\pi}
=\frac{1}{4\pi R
\cosh^2\alpha}\frac{1+\nu}{\sqrt{\nu}}\sqrt{\frac{1-\lambda
}{\lambda(1+\lambda)}}.
\end{eqnarray}
This result is exactly consistent with Hawking temperature derived
by cancelling gauge and gravitational anomalies at the horizon of
the charged black ring (Chen and He's paper in \cite{r16}). Here
to reduce the higher dimensional theory to the effective two
dimensional theory, a dimensional reduction technique is carried
out by using scalar field near the horizon of the charged black
ring. So the resulting Hawking temperature is only for scalar
particles across the horizon. Now we can also conclude that scalar
and Dirac particles can tunnel across the horizon of black rings
at the same Hawking temperature.

\section{Conclusions and discussions} \label{4}

Hawking radiation of scalar particles across black holes or
black rings have been discussed a lot via different methods, such
as recently hot discussing tunnelling method, and anomalous
cancellation method, etc. And Hawking radiation of Dirac particles
across $3$- or $4$-dimensional black holes have also been
presented in recent papers via fermions tunnelling method. In this
paper, choosing a set of appropriate matrices $\gamma^\mu$ for the
the $5$-dimensional neutral, diploe and charged black rings, we
successfully recover Hawking temperatures of these black rings via
fermions tunnelling method.

Fermions tunnelling method has already been successfully applied
to derive Hawking radiation of Dirac particles across stationary
back holes\cite{r9,r10,r12,r13,r14} and black rings (appeared in
this paper). For a non-stationary black hole, although \cite{r11}
has discussed fermions tunnelling from Barddeen-Vaidya and
cosmological black holes, there is no coupling effect between the
spin of Dirac particles and the angular momentum of the black hole
in the tunnelling rate. That is because the involved
non-stationary black holes in \cite{r11} is of spherical symmetry,
and have no angular momentum for themselves. So we expect when
Dirac particles tunnelling from non-stationary black holes with
one or more angular momentum, the spin coupling effect will be
present. This is our next job. In addition, noted that choosing a
set of appropriate matrices $\gamma^\mu$ is a important technique
for fermions tunnelling method, or we can not correctly recover
Hawking temperature we expected. Finally, it is need to say that,
in Sec.(\ref{2}) and (\ref{3}), we only consider the case of Dirac
particles with spin up. In fact, adopting a similar procedure, we
will find the same result for Dirac particles with spin down. That
meas both spin up and spin down Dirac particles tunnel across the
horizon at the same Hawking temperature\cite{r13}, and such
handling does not loss its generality.

\section*{Acknowledgements}
This work is supported by National Natural Science Foundation of
China with Grant Nos. 10675051, 70571027, 10635020, a grant by the
Ministry of Education of China under Grant No 306022, and a
Graduate Innovation Foundation by CCNU.


\end{document}